\documentclass[a4paper]{jpconf}
\usepackage{graphicx,hyperref}
\usepackage{amsmath,amssymb}
\usepackage{iopams}
\usepackage{bm}% bold math
\usepackage{color}

\begin{document}
\title{Photoinduced Phase Transition in Two-Band model on Penrose Tiling}

\author{Ken Inayoshi, Yuta Murakami, and Akihisa Koga}
\address{Department of Physics, Tokyo Institute of Technology, Meguro, Tokyo 152-8551, Japan}

\ead{k-inayoshi@stat.phys.titech.ac.jp}

\begin{abstract}
  We study the effects of the photo irradiation on the band insulating state
  in the two-band Hubbard model on the Penrose tiling.
  Examining the time- and site-dependent physical quantities,
  we find that the excitionic state is dynamically induced
  with site-dependent order parameters.
  It is also clarified that, 
  in the excitonic state induced by the photo irradiation,
  local oscillatory behavior appears in the electron number
  as well as in the order parameter,
  which should be characteristic of the quasiperiodic lattice.
\end{abstract}

Quasicrystal is an interesting system, which is characterized by a long-range order with no translational symmetry~\cite{PhysRevLett.53.1951}.
In the field of quasicrystals, the discovery of an Al-Si-Ru semiconducting approximant~\cite{PhysRevMaterials.3.061601} stimulates
theoretical and experimental investigations of the electron correlation effects on the semiconducting quasicrystals.
In our previous works~\cite{Inayoshi1,Inayoshi2},
we have studied the excitonic insulating phase~\cite{PhysRev.158.462}
on the Penrose tiling~\cite{penroselattice}
to clarify the role of the quasiperiodic structure in the equilibrium and nonequilibrium states.
One of the interesting features is that local charge fluctuations are
induced by the photo irradiation in the BEC-type excitonic insulating state.
Then, a question arises: are similar charge fluctuations induced
even in the photo-excited band insulator on the quasiperiodic lattice? 
It is instructive to treat the band insulating state on the Penrose tiling
to study nonequilibrium behavior after the photo irradiation.

%%%%%%%%%%%%%%%%%%%%%%%%%%%%%%%%%%%%%%%%%%%%%%%%%%%%%%%
\begin{figure}[h]
\begin{center}
\includegraphics[width=0.8\linewidth]{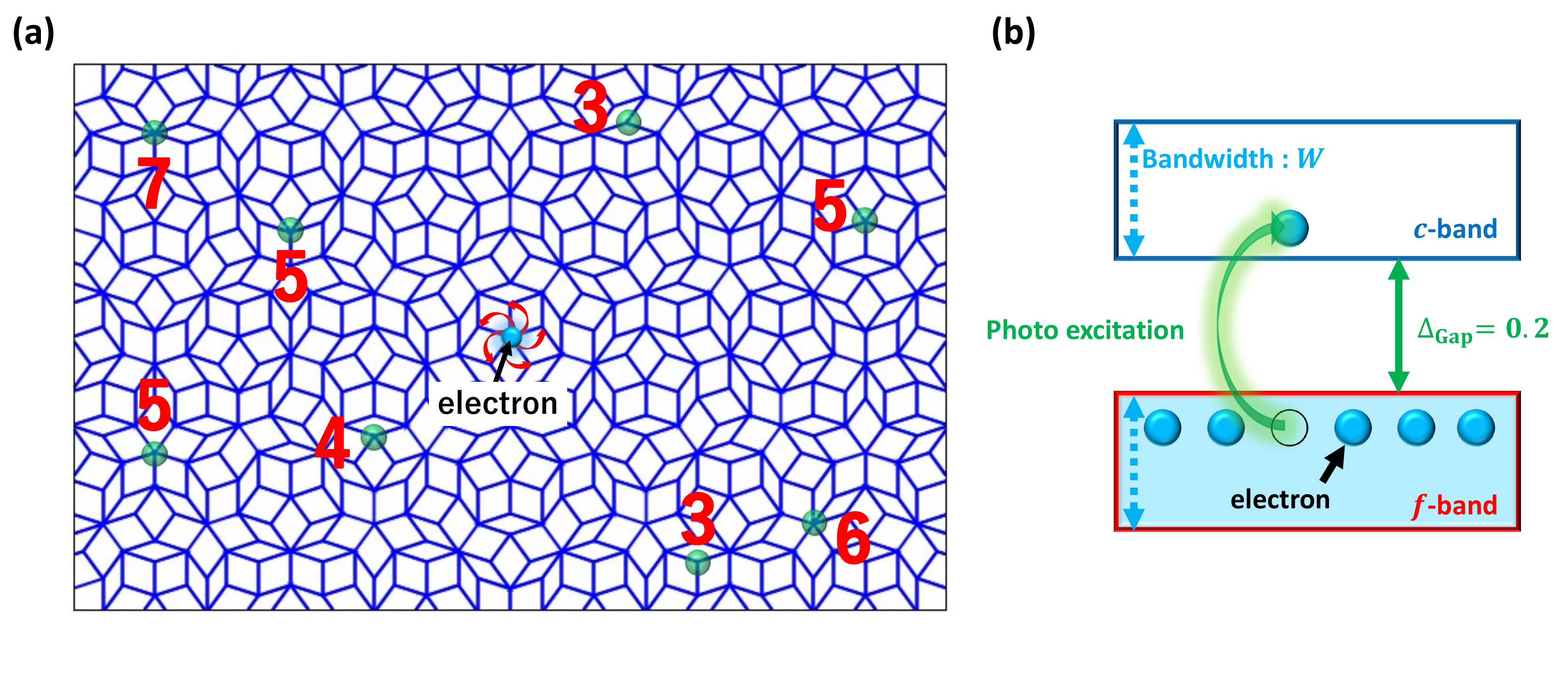}
\caption{
(a) Penrose tiling. The integer for each site represents its coordination number.
(b) Schematic picture of the photo excitation of the band insulating state.
}
\label{setup}
\end{center}
\end{figure}
%%%%%%%%%%%%%%%%%%%%%%%%%%%%%%%%%%%%%%%%%%%%%%%%%%%%%%%

We consider the two-band Hubbard model~\cite{PhysRevB.97.115105,PhysRevB.102.075118}, whose Hamiltonian is given as
\begin{align} \label{eq:hamiltonian}
  \hat{H}_0&=-J\sum_{\langle i,j \rangle \sigma}(\hat{c}_{i\sigma}^{\dagger}\hat{c}_{j\sigma}-\hat{f}_{i\sigma}^{\dagger}\hat{f}_{j\sigma}) 
 +\frac{D}{2}\sum_{i\sigma}(\hat{n}_{ci\sigma}-\hat{n}_{fi\sigma}) 
 -\mu\sum_{i\sigma}(\hat{n}_{fi\sigma} +\hat{n}_{ci\sigma}) \nonumber \\
 &+U\sum_{i}(\hat{n}_{ci\uparrow}\hat{n}_{ci\downarrow}+\hat{n}_{fi\uparrow}\hat{n}_{fi\downarrow})+V\sum_{i\sigma\sigma^{'}}\hat{n}_{ci\sigma}\hat{n}_{fi\sigma'},
\end{align}
where $\hat{a}^\dagger_{i\sigma}$ creates an electron at site $i$
with spin $\sigma\in \{ \uparrow, \downarrow \}$ in the $a(=c, f)$-band,
and $\hat{n}_{ai\sigma}=\hat{a}^\dagger_{i\sigma}\hat{a}_{i\sigma}$.
$J$ is the hopping amplitude between the nearest neighbor sites,
$D$ is the energy difference between two bands, and
$\mu$ is the chemical potential.
$U\ (>0)$ is the intraband onsite interaction
and $V\ (>0)$ is the interband onsite interaction.
We consider here the Penrose tiling as a simple quasiperiodic lattice,
as shown in Fig. ~\ref{setup}(a).
Electronic properties on the Penrose tiling have been discussed
in the tight-binding~\cite{Kohmoto,PhysRevB.38.1621,PhysRevLett.109.106402,PhysRevLett.109.116404,Fan2021},
single-band Hubbard~\cite{Takemori,PhysRevB.96.214402},
and Anderson lattice models~\cite{Takemura}.
One of the important features of the Penrose tiling
is that the number of nearest neighbor sites (coordination number)
$\alpha$ takes 3 to 7 [see Fig.~\ref{setup}(a)].
In this study, we focus on the $\alpha$ dependence of the physical quantities
to discuss the effects of the quasiperiodic tiling.
In the following, our discussions are restricted to the half-filled paramagnetic system,
where $\mu=U/2+V$ and the spin indices are omitted.

To study the photo irradiation in the system,
we introduce the Hamiltonian for the dipole transition as
\begin{align}
  H'(t)&=F(t)\sum_{i}(\hat{c}_{i}^{\dagger}\hat{f}_{i}+{\rm h.c.}),
\end{align}
where $|F(t)|$ is the time dependent amplitude of the external field.
Here, we employ the time-dependent real-space mean-field (MF) approximation.
This method has an advantage in treating the large system,
which allows us to discuss the effect of the quasiperiodic structure.
The electron number of the $a$-band and the local order parameter at the $i$th site
are defined as
$n_{ai}(t)=\langle\psi(t)|\hat{n}_{ai}|\psi(t)\rangle$, 
and $\Delta_{i}(t)=\langle\psi(t)|\hat{c}_{i}^{\dagger}\hat{f}_{i}|\psi(t)\rangle$.
Here $|\psi(t)\rangle=T_{t}{\rm exp}\left[-i\int_{0}^{t}\hat{H}_{\rm MF}^{\rm total}(t') dt'\right]|\psi(0)\rangle$,
where $T_{t}$ is the time-ordering operator and
$|\psi(0)\rangle$ is the ground state of $\hat{H}_{\rm MF}^{\rm total}(t=0)$.
The explicit form of the MF Hamiltonian is given as,
\begin{align} \label{eq:TDMF}
  \hat{H}^{\rm total}_{\rm MF}(t)&=-J\sum_{\langle i,j \rangle}(\hat{c}_{i}^{\dagger}\hat{c}_{j}
  -\hat{f}_{i}^{\dagger}\hat{f}_{j}) 
 +\frac{D}{2}\sum_{i}(\hat{n}_{ci}-\hat{n}_{fi})-\mu\sum_{i}(\hat{n}_{fi} +\hat{n}_{ci}) \nonumber \\
&+U\sum_{i}({n}_{ci}(t)\hat{n}_{ci}+{n}_{fi}(t)\hat{n}_{fi})+2V\sum_{i}({n}_{fi}(t)\hat{n}_{ci}+{n}_{ci}(t)\hat{n}_{fi}) \nonumber \\
&-\sum_{i}\Big[\left\{V\Delta_i(t)-F(t)\right\}\hat{f}_{i}^{\dagger}\hat{c}_{i}+{\rm h.c.}\Big].
\end{align}
We define the single-particle density matrix 
$\rho_{ia,jb}(t)=\langle\psi(t)|\hat{b}_{j}^{\dagger}\hat{a}_{i}|\psi(t)\rangle$.
The time evolution of density matrix is given by
$i\partial_t\bm{\rho}(t)=\left[ \bm{H}^{\rm MF}(t), \bm{\rho}(t) \right]$.
Here, $\bm{H}^{\rm MF}(t)$ is the matrix representation of $\hat{H}^{\rm total}_{\rm MF}(t)$.
Since $\bm{H}^{\rm MF}(t)$ depends on ${\bm{\rho}(t)}$,
one can numerically solve this differential equation.
We note that $\partial_t n_{ci}(t)=-iJ\left\{\sum_{m}\rho_{ic,mc}(t)-\sum_{m}\rho_{mc,ic}(t)\right\}$
if $F(t)=0$,
where $m$ runs the nearest neighbor sites for site $i$.
Therefore, $\sum_i\partial_tn_{ci}(t)=0$.
This implies that the total number of conduction electrons
is conserved when $F(t)=0$ (for example, after the photo irradiation).
However, this does not necessarily imply the absence of local charge fluctuations.
In fact, we have found local charge oscillation triggered by the photo irradiation
in the two-band system in the BEC regime~\cite{Inayoshi2}.
In this study, we examine local physical quantities to clarify whether or not
such charge fluctuations are induced by the photoinduced phase transition
in the band insulating state.

In the following, we take $J$ as the unit of the energy.
For simplicity, we set the parameters as $U=D=4$.
We treat the Penrose tiling with the total sites $N=11006$
under the open boundary condition.
It has been clarified that in the case of $V<V_c \ (\sim 4.3)$, the order parameter is finite and
the excitonic insulating state is realized.
On the other hand, when $V>V_c$,
the large interband interaction stabilizes the band insulating state
with the energy gap $\Delta_{\rm Gap}=2V-W$,
where $W \ (=8.46)$.
In this study, to discuss the effects of the photo irradiation on the band insulating state,
we set the interband interaction as $V=4.33$, where $\Delta_{\rm Gap}=0.2$, see Fig.~\ref{setup}(b).
Then, we apply the single-cycle pulse field to the two-band system,
which is given as $F(t)=F_0\sin\omega t \; \theta(t)\theta(t_p-t)$ with
the magnitude of the external field $|F_0|$, the frequency $\omega$,
the Heaviside step function $\theta(t)$, and
the light irradiation time $t_p=2\pi/\omega$~\cite{PhysRevB.97.115105}.
Setting $\omega=2\Delta_{\rm Gap}$, we study the coordination number dependent
quantities $\Delta_\alpha(t)=\sum_{i\ \text{with}\ Z_i=\alpha}|\Delta_i(t)|/N'_\alpha$ and $n_c^\alpha(t)=\sum_{i\ \text{with}\ Z_i=\alpha}n_{ci}(t)/N'_\alpha$ to discuss the photoinduced phase transition
in the two-band system on the quasiperiodic tiling.
Here, $N'_\alpha$ is the number of the lattice sites with $Z_i=\alpha\ (=3,\cdots,7)$ in the bulk region.

%%%%%%%%%%%%%%%%%%%%%%%%%%%%%%%%%%%%%%%%%%%%%%%%%%%%%%%
\begin{figure}
\begin{center}
\includegraphics[width=0.99\linewidth]{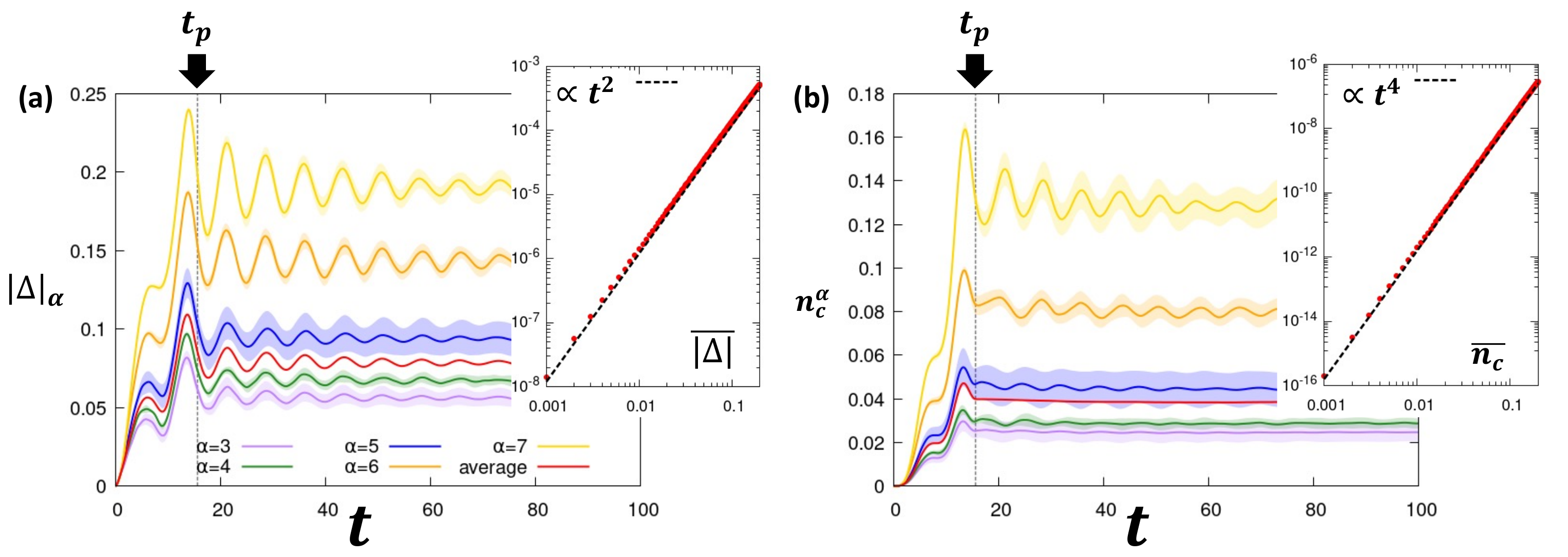}
\caption{
  Time evolution of (a) $|\Delta|_\alpha$ and (b) $n_c^\alpha$ for each $\alpha$
  after the single-cycle pulse is injected with $F_0=-0.07$.
  We also plot the bulk averaged values, $\overline{|\Delta|}$ and $\overline{n_c}$.
  Note that we define those averaged values in the {\it bulk} region which excludes edge sites.
  Therefore, $\overline{n_c}$ is not conserved (slightly oscillates) 
  although the total number of conduction electrons is conserved at $t>t_p$.
  Right panels show the log-log plots of $\overline{|\Delta|}$ and $\overline{n_c}$.
  }
\label{each}
\end{center}
\end{figure}
%%%%%%%%%%%%%%%%%%%%%%%%%%%%%%%%%%%%%%%%%%%%%%%%%%%%%%%

When $t<0$, no field is applied and 
the band insulating state with $\Delta_{\rm Gap}=0.2$ is realized with $\Delta_i=0$ and $n_{ci}=0$.
Switching on the photo irradiation at $t=0$,
local order parameter $|\Delta_i|$ and its bulk average $\overline{|\Delta|}$ is induced,
as shown in Fig.~\ref{each}(a).
We also find in Fig.~\ref{each}(b) that the bulk averaged value $\overline{n_c}$ becomes finite,
implying that some conduction electrons are excited.
Our MF calculations clearly show $\overline{|\Delta|}\sim t^2$ and $\overline{n_c}\sim t^4$, 
which means that the photoinduced phase transition occurs at $t=0$.
We also find that 
$\Delta_\alpha(t)$ and $n_c^\alpha(t)$ are well classified by the coordination number,
as seen in Figs.~\ref{each}(a) and (b).
We wish to note that when $t>t_p$, the total number of conduction electrons is constant, while $n_c^\alpha(t)$ shows oscillatory behavior.
This implies that local charge fluctuations are induced even in the band insulating state.
This is similar to the behavior in the BEC-type excitonic state in the two-band model on the Penrose tiling~\cite{Inayoshi2}.
Therefore, the charge fluctuations are characteristic nonequilibrium phenomena peculiar to the strong coupling regime of quasicrystalline excitonic state.

To summarize, we have studied the effects of the photo irradiation on the band insulating state
in the two-band Hubbard model on the Penrose tiling
by means of the time- and site-dependent MF approximation.
It is found that the photoinduced phase transition to the excitonic state occurs in the quasiperiodic lattice as in normal crystal~\cite{ostreich1993,PhysRevMaterials.3.124601,PhysRevB.101.035203}.
In the photo-induced state, the interband interaction is so large that 
the electron-hole pairs are tightly coupled and these are almost localized around each lattice site.
Therefore, physical quantities are well classified by the coordination number.
In general, each site in a quasiperiodic tiling is topologically inequivalent.
It is expected that this unusual charge oscillation should appear
in the realistic quasicrystals and approximants,
which will be discussed in the future.

\ack
K.I. acknowledges the financial support from 
Advanced Human Resource Development Fellowship for Doctoral Students, Tokyo Institute of Technology.
This work was supported by a Grant-in-Aid for Scientific Research from JSPS,
KAKENHI Grant Nos. JP20K14412, JP20H05265, JP21H05017 (Y.M.),
JP21H01025, JP19H05821, JP18K04678, JP17K05536 (A.K.),
JST CREST Grant No. JPMJCR1901 (Y.M.).
\section*{References}
\bibliographystyle{iopart-num}
\bibliography{refs}

\end{document}